%% file: main.tex
\documentclass{article}
\usepackage{spconf,amsmath,graphicx,hyperref,amssymb,amsthm,mathtools,cleveref,color,subcaption,enumerate}

\title{A simple bound on the BER of the MAP decoder\\ for massive MIMO systems}
\name{Christos Thrampoulidis$^\star$, Ilias Zadik$^\dagger$
, Yury Polyanskiy$^\dagger$
\thanks{
This material is based upon work supported by the National Science
   Foundation under Grant No CCF-17-17842.}
\address{$^\star$ University of California, Santa Barbara, USA. \\
$^\dagger$ Massachusetts Institute of Technology, USA.}
}

\input{commands}

\begin{document}

%
\ninept

\maketitle
\begin{abstract}
The deployment of massive MIMO systems has revived much of the interest in the study of the large-system performance of multiuser detection systems. In this paper, we prove a non-trivial upper bound on the bit-error rate (BER) of the MAP detector for BPSK signal transmission and equal-power condition. In particular, our bound is approximately tight at high-SNR. The proof is simple and relies on Gordon's comparison inequality. Interestingly, we show that under the assumption that Gordon's inequality is tight, the resulting BER prediction matches that of the replica method under the replica symmetry (RS) ansatz. Also, we prove that, when the ratio of receive to transmit antennas exceeds $\deltazero$, the replica prediction matches the matched filter lower bound (MFB) at high-SNR. We corroborate our results by numerical evidence. 
\end{abstract}
 \begin{keywords}
massive mimo, large-system analysis, JO detector, Gaussian process inequalities, replica method.
\end{keywords}


\input{intro}
\input{upperBound}

\input{proof}
\input{replica}

\input{conclusion}

\section*{Acknowledgement}
The authors would like to thank Dr. Tom Richardson for sharing with us his proof of Proposition \ref{lem:ell_critical}. In particular, Proposition \ref{lem:ell_critical} as it appears here is a refined version of what appeared in our initial submission.

\input{appendix}

\input{fullVer_appendix}

\bibliography{compbib}


\end{document}

%% file: commands.tex
\newcommand{\ellt}{\tilde{\ell}}

\newcommand{\deltazero}{0.9251}
\newcommand{\sigmazero}{0.1419}

\newcommand{\down}[1]{{#1}^{\downarrow}}

\newcommand{\thetastar}{\theta_\star}
\newcommand{\thetazero}{\theta_0}
\newcommand{\alphazero}{\alpha_0}

\newcommand{\eqP}{\stackrel{P}{=}}
\newcommand{\leqP}{\stackrel{P}{\leq}}

\newcommand{\BER}{\mathrm{BER}}

\usepackage{dsfont}
\newcommand{\ind}[1]{{\mathds{1}}_{\{#1\}}}

\newcommand{\simiid}{\stackrel{\text{iid}}{\sim}}

\theoremstyle{theorem}
\newtheorem{propo}{Proposition}[section]

\newtheorem{thm}{Theorem}[section]

\theoremstyle{remark}
\newtheorem{remark}{Remark}

\theoremstyle{definition}

\newcommand{\eps}{\epsilon}

\newcommand{\sign}{\mathrm{sign}}

\newcommand{\SNR}{\mathrm{SNR}}

\newcommand{\E}{\mathbb{E}}                    

\newcommand{\sig}{\sigma}

\newcommand{\nn}{\notag}
\newcommand{\R}{\mathbb{R}}

\newcommand{\G}{\mathbf{G}}

\newcommand{\A}{\mathbf{A}}

\newcommand{\x}{\mathbf{x}}
\newcommand{\w}{\mathbf{w}}

\newcommand{\g}{\mathbf{g}}

\newcommand{\y}{\mathbf{y}}

\newcommand{\z}{\mathbf{z}}

\newcommand{\ab}{\mathbf{a}}

\newcommand{\h}{\mathbf{h}}

\newcommand{\Sc}{{\mathcal{S}}}

\newcommand{\Nn}{\mathcal{N}}

\newcommand{\Ec}{\mathcal{E}}

\newcommand{\beq}{\begin{equation}}
\newcommand{\eeq}{\end{equation}}
\newcommand{\bea}{\begin{align}}
\newcommand{\eea}{\end{align}}



%% file: intro.tex
\section{Introduction}
Massive multiple-input multiple-output (MIMO)  systems, where the base station is equipped with hundreds of thousands of antennas, promise improved spectral efficiency, coverage and range compared to small-scale systems. As such, they are widely believed to play an important role in  5G wireless communication systems \cite{larsson2014massive}. Their deployment has revived much of the recent interest for the study of multiuser detection schemes in high-dimensions, e.g., \cite{ngo2013energy,wen2014message,narasimhan2014channel,MIMO_maleki}.  

A large host of exact and heuristic detection schemes have been proposed over the years.  
Decoders such as  zero-forcing (ZF) and linear minimum mean square error (LMMSE) have  inferior performances \cite{verdu1998multiuser}, and others such as local neighborhood search-based methods \cite{vardhan2008low} and lattice reduction-aided (LRA) decoders \cite{windpassinger2003low,zhou2013element} are often difficult to precisely characterize. Recently, \cite{TSP18} studied in detail the performance of the box-relaxation optimization (BRO), which is a natural convex relaxation of the maximum a posteriori (MAP) decoder, and which allows one to recover the signal via efficient convex optimization followed by hard thresholding. In particular, \cite{TSP18} precisely quantifies the performance gain of the BRO compared to the ZF and the LMMSE. Despite such gains, it remains unclear the degree of sub-optimality of the convex relaxation compared to the combinatorial MAP detector. The challenge lies in the complexity of analyzing the latter. In particular, known predictions of the performance of the MAP detector are known only via the (non-rigorous) replica method from statistical physics \cite{Tanaka}. 

In this paper, we derive a simple, yet non-trivial, upper bound on the bit error rate (BER) of the MAP detector. We show (in a precise manner) that our bound is approximately tight at high-SNR, since it is close to the matched filter lower bound (MFB). Our numerical simulations verify our claims and further include comparisons to the replica prediction and to the BER of the BRO. Our proof relies on Gordon's Gaussian comparison inequality \cite{Gor88}. While Gordon's inequality is not guaranteed to be tight, we make the following possibly interesting and useful observation. If Gordon's inequality was asymptotically tight, then its BER prediction would match the prediction of the replica method (under replica-symmetry). 

\vspace{-10pt}
\section{Setting}
\vspace{-5pt}
We assume a real Gaussian wireless channel, additive Gaussian noise and and uncoded modulation scheme. For concreteness, we focus on the binary-phase-shift-keying (BPSK) transmission; but, the techniques naturally extend to other constellations.  Formally, we seek to recover an $n$-dimensional BPSK vector $\x_0\in\{\pm1\}^n$ from the noisy MIMO relation
$\y=\A\x_0+\sigma \z \in\R^m,$
where $\A\in\R^{m\times n}$ is the channel matrix (assumed to be known) with entries iid $\Nn(0,1/n)$. and $\z\in\R^m$ the noise vector with entries iid $\Nn(0,1)$. The normalization is such that the reciprocal of the noise variance $\sigma^2$ is equal to the SNR, i.e.,
$
\SNR = 1/{\sigma^2}.
$
The performance metric of interest is the bit-error rate (BER). The BER of a detector which outputs $\hat\x$ as an estimate to $\x_0$ is formally defined as $\BER :=  \frac{1}{n}\sum_{i=1}^n \ind{\hat\x_i\neq \x_{0,i}}.$

In this paper, we study the BER of the MAP (also commonly referred to in this context as the jointly-optimal (JO) multiuser) detector, which is defined by
\begin{align}
\hat\x = \arg\min_{ \substack{\x\in\{\pm1\}^n}}\|\y-\A\x\|_2.\label{eq:MAP}
\end{align}
We state our results in the large-system limit where $m,n\rightarrow\infty$, while the ratio of receive to transmit antennas is maintained fix to $\delta=m/n$\footnote{The proof of our main result Theorem \ref{thm:main} reveals that a non-asymptotic bound is also possible with only slight more effort.}
It is well known that in the worst case, solving \eqref{eq:MAP} is an NP-hard combinatorial optimization problem in the number of users \cite{verdu1989computational}. 
The BRO is a relaxation of \eqref{eq:MAP} to an efficient convex quadratic program, namely
$
\hat\x=\sign\big(\arg\min_{\x\in[-1,1]^n}\|\y-\A\x\|_2\big).
$
Its performance in the large-system limit has been recently analyzed in \cite{TSP18}. Regarding the performance of \eqref{eq:MAP}, Tse and Verdu \cite{tse2000optimum} have shown that the BER approaches zero at high-SNR. Beyond that, there is a now long literature that studied \eqref{eq:MAP} using the replica method, developed in the field of spin-glasses.  The use of the method in the context of multiuser detection was pioneered by Tanaka \cite{Tanaka} and several extensions have followed up since then \cite{guo2005randomly,caire2004iterative}. The replica method has the remarkable ability to yield highly nontrivial predictions, which in certain problem instances they can been formally shown to be correct (e.g., \cite{talagrand2003spin,reeves2016replica,barbier2016mutual}). However, it is still  lacking a complete rigorous mathematical  justification. 

%% file: upperBound.tex
\section{Results}

\subsection{Upper bound}
This section contains our main result: a simple  upper bound on the BER of \eqref{eq:MAP}.
First, we introduce some useful notation. We say that an event $\Ec(n)$ holds with probability approaching 1 (wpa 1) if $\lim_{n\rightarrow}\Pr(\Ec(n))=1$. Let $X_n$ a sequence of random variables indexed by $n$ and $X$  some constant. We write $X_n\eqP X$ and $X_n\leqP X$, if for all $\eps>0$ the events $\{|X_n-X|\leq \eps\}$ and $\{X_n\leq X+\eps\}$ hold wpa 1. Finally, let $\phi(x)=\frac{1}{\sqrt{2\pi}}e^{-x^2/2}$, $Q(x)=\int_{x}^{\infty}\phi(\tau)\mathrm{d}\tau$ the Gaussian tail function and $Q^{-1}$ its inverse.

\begin{thm}\label{thm:main}
Fix constant noise variance $\sig^2>0$ and $\delta>0$. Let $\BER$ denote the bit-error-rate of the MAP detector in \eqref{eq:MAP} for fixed but unknown $\x_0\in\{\pm1\}^n$. Define the function $\ell(\theta):(0,1)\rightarrow\R$:
\begin{align}\label{eq:ell}
\ell(\theta):= \sqrt{\delta}\,\sqrt{4\theta+\sig^2}-\sqrt{\frac{2}{\pi}}e^{-\frac{(Q^{-1}(\theta))^2}{2}},
\end{align}
and let $\thetazero\in(0,1)$ be the largest solution to the equation $\ell(\theta)=\sig\sqrt{\delta}$.
Then, in the limit of $m,n\rightarrow\infty$, $\frac{m}{n}=\delta$, it holds 
$
\BER \leqP \thetazero.
$
\end{thm}

Propositions \ref{lem:ell_critical} and \ref{lem:ell_more} in the Appendix gathers several useful properties of the function $\ell$. 
Notice that $\ell(1^-)>\ell(0^+)=\sqrt{\delta}\sigma$. Also, $\ell$ is continuous and  $\ell'(0^+)<0$. Thus, $\thetazero$ in Theorem \ref{thm:main} is well-defined. Moreover, we show in Proposition \ref{lem:ell_more}(i) that if $\delta>\deltazero$ or $\sigma^2>\sigmazero$, then $\thetazero$ is the \emph{unique} solution of the equation $\ell(\theta)=\sigma\sqrt{\delta}$ in $(0,1)$.

\begin{remark}[On the function $\ell(\theta)$]\label{rem:ell}
Let us elaborate on the operational role of the function $\ell$. We partition the feasible vectors $\x\in\{\pm 1\}^n$ according to their Hamming distance from the true vector $\x_0$. Specifically, for $\theta\in[0,1]$ let $\Sc_{\theta}:=\{\x\in\{\pm 1\}^n\,:\,\|\x-\x_0\|_0=\theta n \}$ and consider the optimal cost of \eqref{eq:MAP} for each partition, i.e., 
\begin{align}\label{eq:cstar}
c_\star({\theta}):=\min_{\x\in\Sc_\theta}\frac{1}{\sqrt{n}}\|\y-\A\x\|_2.
\end{align}
 Evaluating the BER of \eqref{eq:MAP} is of course closely related to understanding the typical behavior of $c_\star(\theta)$ in the large system limit. The proof of the theorem in Section \ref{sec:proof} shows that $\ell(\theta)$ is a \emph{high-probability lower bound} on $c_\star(\theta)$. Hence, we get an estimate on the BER via studying $\ell(\theta)$ instead. In this direction, note that the value $\sigma\sqrt{\delta}$, to which $\ell(\theta)$ is compared to, is nothing but the typical value of $c_\star(0)=\frac{\|\y-\A\x_0\|_2}{\sqrt{n}}=\frac{\|\z\|_2}{\sqrt{n}}$. Finally, we make the following note for later reference: the value $\inf_{\theta\in(0,1)} \ell(\theta)$ is a high-probability lower bound to the optimal cost of \eqref{eq:MAP}, i.e., to $c_\star=\inf_{\theta\in(0,1)}c_\star(\theta)$. An illustration of these is included in Figure \ref{fig:sketch}.
%
%
%
\end{remark}

\begin{remark}[A genie lower bound]\label{rem:MFB}
A lower bound on the BER of \eqref{eq:MAP} can be obtained easily via comparison to the idealistic matched filter bound (MFB), where one assumes that all $n-1$, but 1, bits of $\x_0$ are known. In particular, the MFB corresponds to the probability of error in detecting (say) $\x_{0,1}\in\{\pm 1\}$ from $\widetilde{\y}=\x_{0,1}\ab_1+\z$, where $\widetilde{\y}=\y-\sum_{i=2}^n \x_{0,i}\ab_i$ is assumed known, and $\ab_i$ is the $i^{\text{th}}$ column of $\A$ (eqv., the MFB is the error probability of an isolated transmission of only the first bit over the channel). It can be shown (e.g., \cite{TSP18}) that the MFB is given by $Q(\sqrt{\delta\,\SNR})$. Combining this with (a straightforward re-parametrization of) Theorem \ref{thm:main} it follows that the BER of \eqref{eq:MAP} satisfies 
\begin{align}\label{eq:MFB}
Q(\sqrt{\delta\,\SNR})\leq \BER\leq Q(\tau_0),
\end{align}
where $\tau_0\in\R$ is the \emph{smallest} solution to the equation
$
\sqrt{\delta \SNR}+2\phi(\tau)=\sqrt{\delta \SNR}\,\sqrt{1+4\SNR\,Q(\tau)} .
$

\begin{remark}[Behavior at high-SNR]\label{rem:high}
In Proposition \ref{lem:ell_more}(iv) we prove that at high values of $\SNR\gg 1$: $\thetazero\rightarrow0$. Thus, from Theorem \ref{thm:main} we have that $\BER$ approaches zero (thus, providing an alternative proof to the corresponding result in \cite{tse2000optimum}). This thinking confirms already that our upper bound is non-trivial. In fact, an even stronger statement can be shown, namely, at $\SNR\gg 1$: $\thetazero\approx Q(\sqrt{\delta\,\SNR}-\eta)$ for an arbitrarily small $\eta>0$ (see Proposition \ref{lem:ell_more}(iv) for exact statement). This, when combined with the MFB in \eqref{eq:MFB} shows that our upper bound is approximately tight at high-SNR.

\end{remark}

%

\end{remark}

\begin{figure}[t]
    \begin{subfigure}[b]{0.5\columnwidth}
        \includegraphics[width=\columnwidth]{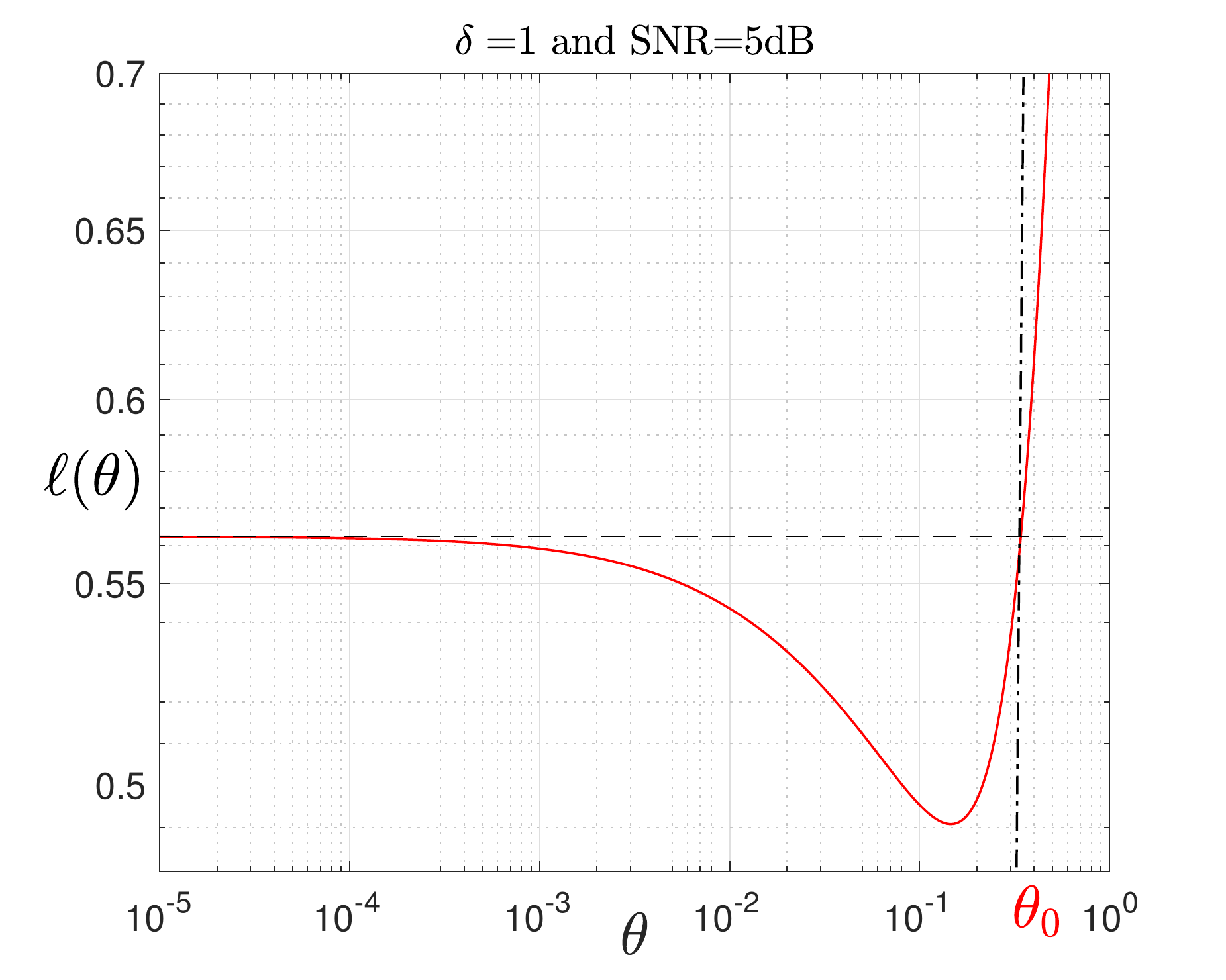}
    \end{subfigure}
    \begin{subfigure}[b]{0.5\columnwidth}
        \includegraphics[width=\columnwidth]{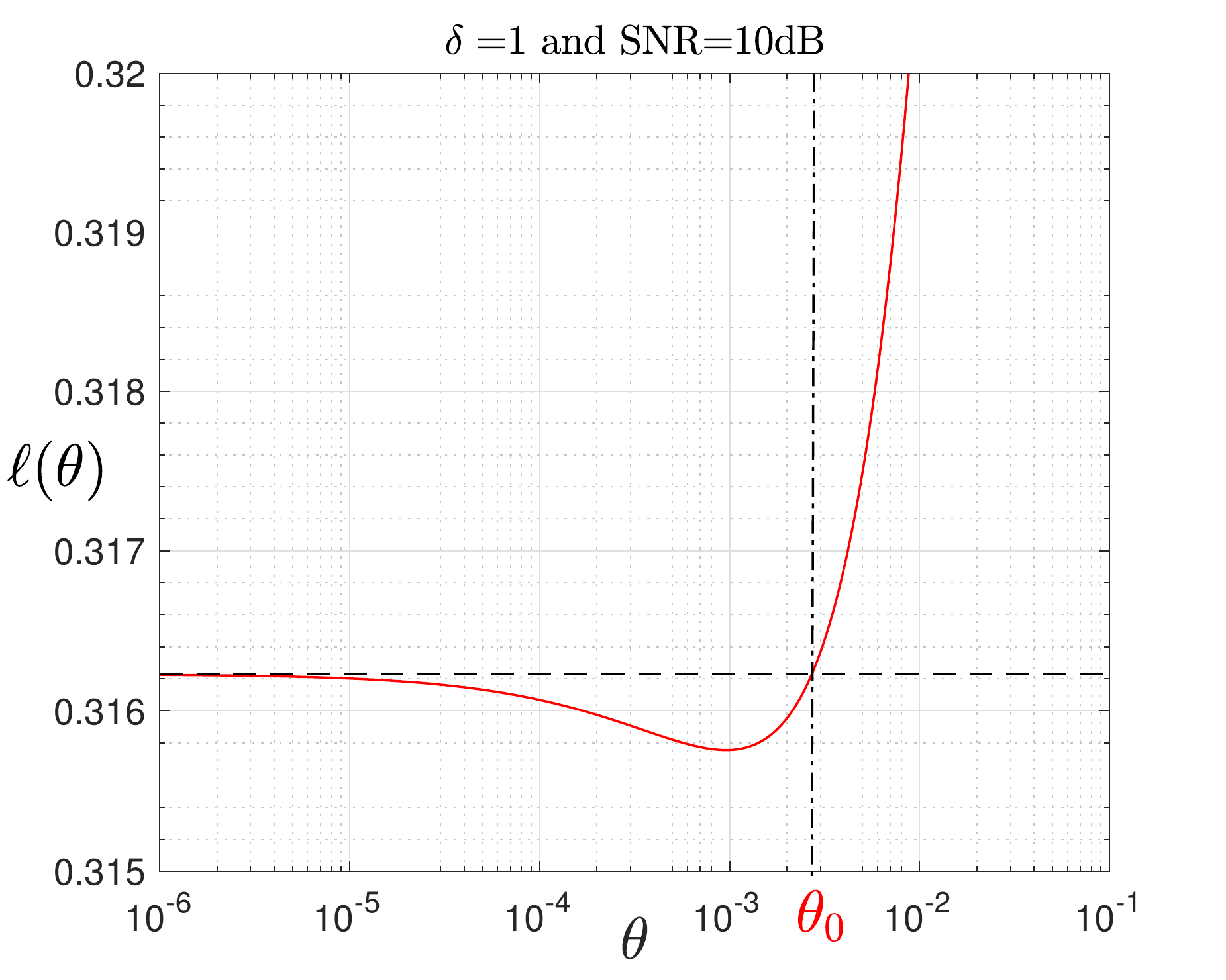}
    \end{subfigure}
\caption{Plots of the function $\ell(\theta)$ defined in \eqref{eq:ell} for two problem instances: ($\delta=1, \SNR=5$dB), ($\delta=1, \SNR=10$dB). Also depicted the value of $\thetazero$ for each instance (see Theorem \ref{thm:main}).}
\label{fig:ell}
\end{figure}

\begin{figure}[t]
\includegraphics[width=.85\columnwidth]{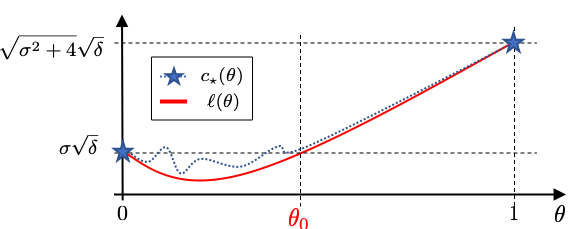}
\caption{The function $\ell(\theta)$ (in red) is a high-probability lower bound on the typical value of $c_\star(\theta)$ (in dashed blue) defined in \eqref{eq:cstar}. See Remark \ref{rem:ell}. }\label{fig:sketch}
\end{figure}

\begin{remark}[Gordon's comparison inequality]\label{rem:GMT}
The proof of Theorem \ref{thm:main} uses Gordon's comparison inequality for Gaussian processes (also known as the Gaussian min-max Theorem (GMT)). In essence, the GMT provides a simple lower bound on the typical value of $c_\star(\theta)$ in \eqref{eq:cstar} in the large-system limit. Gordon's inequality is classically used to establish (non)-asymptotic probabilistic
lower bounds on the minimum singular value of Gaussian matrices \cite{davidson2001local}, and
has a number of other applications in high-dimensional convex geometry \cite{Ledoux}. In general, the inequality is not tight. Recently, Stojnic \cite{StoLASSO}  proved that the inequality is tight when applied to convex problems. The result was refined in \cite{COLT15} and has been successfully exploited to precisely analyze the BER of the BRO \cite{TSP18}. Unfortunately, the minimization in \eqref{eq:cstar} is not convex, thus there are no immediate tightness guarantees regarding the lower bound $\ell(\theta)$. Interestingly, in Section \ref{sec:replica} we show that if GMT was (asymptotically) tight then it would result in a prediction that matches the replica prediction in \cite{TanakaNIPS}.
\end{remark}

\begin{remark}[Replica prediction]\label{rem:replica}
The replica prediction on the BER of \eqref{eq:MAP} is given by \cite{Tanaka} (based on the ansatz of replica-symmetry (RS)) as the solution to a system of nonlinear equations. It is reported in \cite[Eqn.~(15)]{TanakaNIPS} that as long as $\delta$ is not too small, the saddle-point equations reduce to the solution of the 
%
following fixed-point equation\footnote{For the reader's convenience we note the following mapping between notation here and \cite{TanakaNIPS}: $\delta\leftrightarrow \alpha$, $\sigma^2\leftrightarrow\beta_s^{-1}$ and $\BER\leftrightarrow (1-m)/2$. }:
\begin{align}\label{eq:replica}
\theta = Q\big(\sqrt{\frac{\delta}{\sig^2+4\theta}}\big).
\end{align}
Onwards, we refer to \eqref{eq:replica} as the replica-symmetry prediction. 
In Proposition \ref{lem:ell_critical}, we prove that equation \eqref{eq:replica} has either one or three solutions. In the later case, the BER formula can exhibit complicated behavior, such as anomalous, non-monotonic dependence on the SNR \cite{Tanaka}. On the other hand, the solution is unique if either $\delta>\deltazero$ or $\sigma^2\geq\sigmazero$. This proves the numerical observations reported in \cite[Fig.~3]{tanaka2001statistical}. Finally, Proposition \ref{lem:ell_more}(iii) shows that when $\delta>\deltazero$ and $\SNR\gg 1$, the unique solution  of \eqref{eq:replica} satisfies $\thetastar\approx Q(\sqrt{\delta\,\SNR})$. This suggests that at high-SNR, the BER of the \eqref{eq:MAP} decreases at an optimal rate. 
\end{remark}


\subsection{Numerical Evaluations}\label{sec:sim}

%
\begin{figure}[t!]
    \centering
    \begin{subfigure}[b]{.9\columnwidth}
        \includegraphics[width=.9\columnwidth]{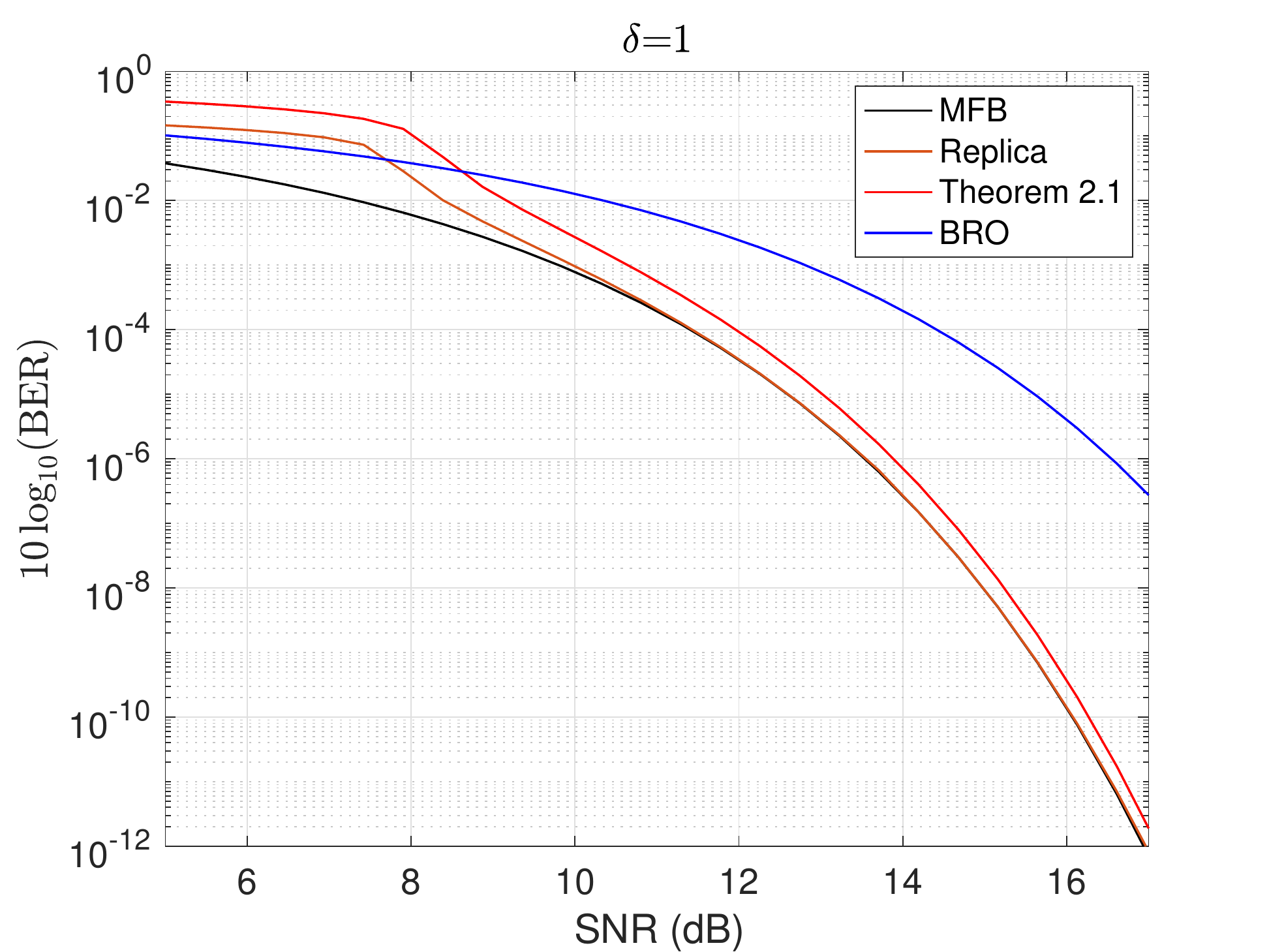}
        \caption{$\delta=1$}
                \label{fig:delta=1}
    \end{subfigure}
    \begin{subfigure}[b]{.9\columnwidth}
        \includegraphics[width=.9\columnwidth]{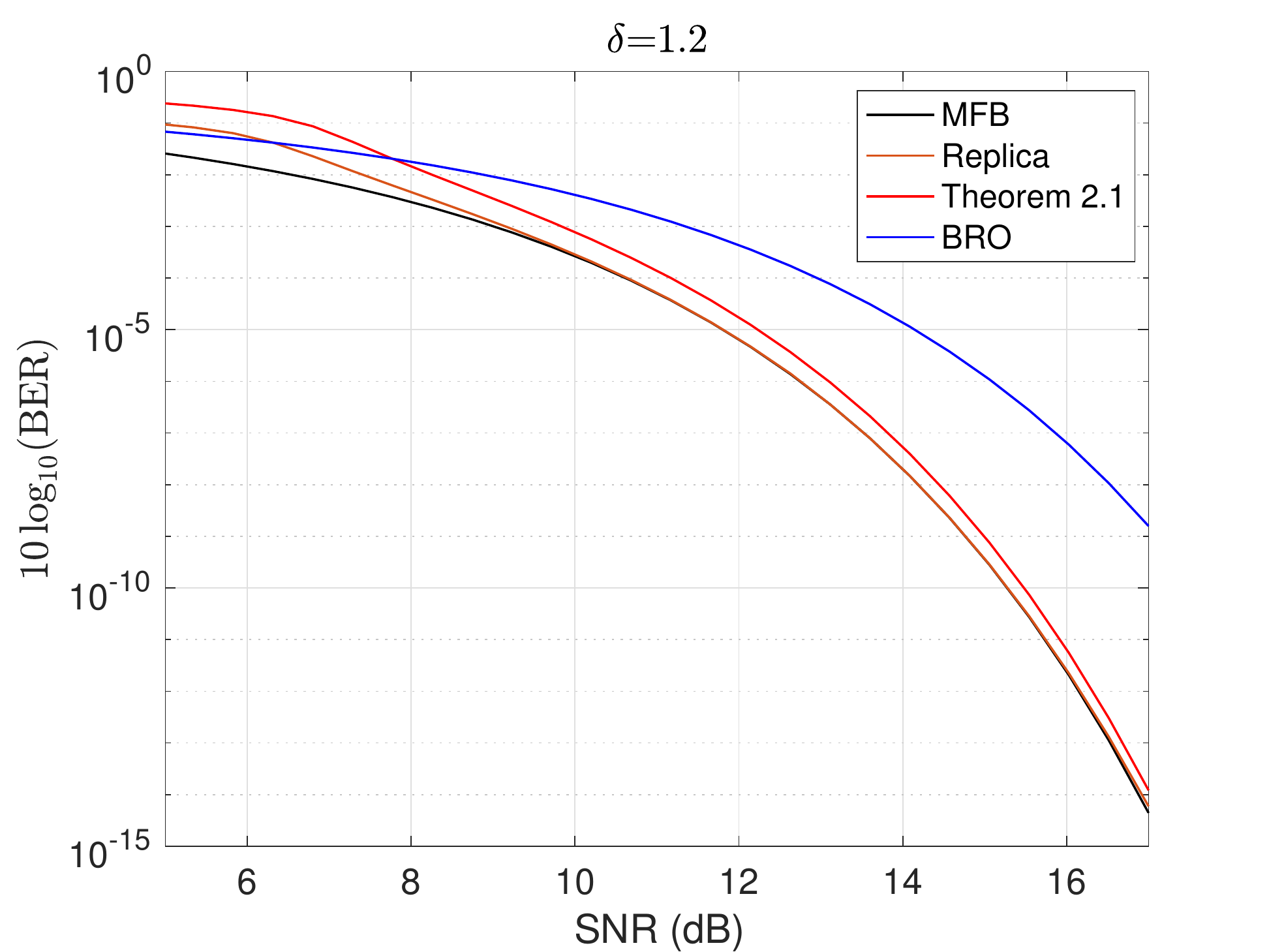}
        \caption{$\delta=1.2$}
        \label{fig:delta=1.2}
    \end{subfigure}
    \caption{BER curve as a function of the SNR (in dB) for the following:  matched-filter lower bound (MFB) (cf. Remark \ref{rem:MFB}); replica prediction corresponding to \eqref{eq:replica}); upper bound of Theorem \ref{thm:main} for \eqref{eq:MAP}; box-relaxation optimization (BRO) \cite{TSP18}.}
    \label{fig:BER_vs_SNR}
\end{figure}

Figure \ref{fig:BER_vs_SNR} includes numerical illustrations that help visualize the prediction of Theorem \ref{thm:main} and several of the remarks that followed. For two values of $\delta$, we plot BER as a function of $\SNR=1/\sigma^2$. Each plot includes four curves: (i) the MFB; (ii) the solution to \eqref{eq:replica} corresponding to the replica prediction; (iii) the upper bound $\thetazero$ of Theorem \ref{thm:main}; (iv) the BER of the BRO according to \cite[Thm.~II.I]{TSP18}. We make several observations. First, it is interesting to note that our upper bound follows the same trend as the replica prediction. For example, note the kink at values of $\SNR\sim 7$dB in both the curves in Figure \ref{fig:delta=1}. Second, note that the upper bound of Theorem \ref{thm:main} approaches the MFB at high-SNR confirming our theoretical findings in Remark \ref{rem:high}. Also, as predicted in Remark \ref{rem:replica}, the solution $\thetastar$ to \eqref{eq:replica} goes to zero exactly at the rate of the MFB. Finally, let us compare the upper bound $\thetazero$ of Theorem \ref{thm:main} to the BER of the BRO. At low SNR, $\thetazero$ takes values larger than the latter. We remark that Theorem \ref{thm:main} is not entirely to be blamed for this behavior, since the replica prediction experiences the very same one. There is no contradiction here:  the MAP detector is \emph{not} optimal for minimizing the BER (e.g., \cite[Sec.~2]{Tanaka}), thus it is likely that its convex approximation (aka, the BRO) shows better BER performance at low-SNR. On the other hand, for high-SNR our upper bound takes values significantly smaller than the BER of the BRO. This proves that at high-SNR the latter is still quite far from that of the combinatorial optimization it tries to approximate.
%

%% file: proof.tex
\subsection{Proof Theorem \ref{thm:main}}\label{sec:proof}
Let $\hat\x$ be the solution to \eqref{eq:MAP}. First, observe that $\|\hat\x-\x_0\|_2^2 = 2n - 2(n - \sum_{i=1}^n \ind{\hat\x_i\neq \x_{0,i}})=4n\,\BER$. Hence, we will prove that 
\begin{align}\label{eq:toshow}
\frac{\|\hat\x-\x_0\|_2}{\sqrt{n}}\leqP \alphazero=:2\sqrt{\thetazero}\in(0,1).
\end{align}
Second, due to rotational invariance of the Gaussian measure we can assume without loss of generality that $\x_0=+\mathbf{1}$. For convenience, define the (normalized) error vector $\w:=n^{-1/2}(\x-\mathbf{1})$ and consider the set of feasible such vectors that do not satisfy \eqref{eq:toshow}, i.e., $$\Sc(\alphazero):=\big\{\w\in\{-2/\sqrt{n},0\}^n~:~\|\w\|_2\geq \alphazero+\eps\,\big\},$$
for some fixed (but arbitrary) $\eps>0$. Also, denote the (normalized) objective function of \eqref{eq:MAP} as $F(\w)=F(\w;\z,\G):=n^{-1/2}\|\z-\G\w\|_2$, where $\G=\sqrt{n}\A$ has entries iid  standard normal. With this notation, our goal towards establishing \eqref{eq:toshow} is proving that there exists constant $\eta:=\eta(\eps)>0$ such that the following holds wpa 1,
\begin{align}\label{eq:toshow2}
\min_{\w\in\Sc(\alphazero)} F(\w) \geq \min_{\w\in\{-2/\sqrt{n},0\}^n} F(\w)+\eta.
\end{align}
Our strategy in showing the above is  as follows. 

First, we use Gordon's inequality to obtain a high-probability lower bound on the left-hand side (LHS) of \eqref{eq:toshow2}. In particular, it can be shown (see for example \cite[Sec.~D.3]{TSP18}) that the primary optimization (PO) in the (LHS) of \eqref{eq:toshow2} can be lower bounded with high-probability by an auxiliary optimization (AO) problem, which is defined as follows:
\begin{align}\label{eq:Gor}
\min_{\w\in\Sc(\alphazero)} G(\w;\g,\h) := \sqrt{\|\w\|_2^2+\sigma^2}\,\|\g\|_2 - \h^T\w,
\end{align}
where $\g\in\R^{m}$ and $\h\in\R^{n}$ have entries iid Gaussian $\Nn(0,1/n)$. Specifically, the following statement holds for all $c\in\R$:
\begin{align}\label{eq:GMT_proof}
\Pr\big(\, \min_{\w\in\Sc(\alphazero)} F(\w;\z,\G) \leq c \,\big) \leq 2\,\Pr\big(\, \min_{\w\in\Sc(\alphazero)} G(\w;\g,\h) \leq c \,\big).
\end{align}

 The AO can be easily simplified as follows
\begin{align}\label{eq:Gor}
\min_{1\geq\alpha\geq\alphazero+\eps}  \sqrt{\alpha^2+\sigma^2}\,\|\g\|_2 - \frac{2}{\sqrt{n}}\sum_{i=1}^{(\alpha^2/4) n}\down{\h}_i ,
\end{align} 
where, $\down{\h}_1\geq\down{\h}_2\geq\ldots\geq\down{\h}_n$ denotes the ordered statistics of the entries of $\h$ and we have used the fact that for $\w\in\{-2/\sqrt{n},0\}^n$ it holds $\|\w\|_2=\alpha \Leftrightarrow\|\w\|_0=\alpha^2/4$. Furthermore, note that $\|\g\|_2\eqP \sqrt{\delta}$ and \footnote{Let $\gamma_i\simiid\Nn(0,1)$ and $\theta\in(0,1)$. Then, for large $n$: $\frac{1}{n}\sum_{i=1}^{\theta n} \down{\gamma}_i \approx \frac{1}{n}\sum_{\{i:\gamma_i\geq Q^{-1}(\theta)\}} h_i \approx \E[\gamma\,|\,\gamma\geq Q^{-1}(\theta)] =  \phi\big( (Q^{-1}(\theta)\big).$} for any fixed $\theta\in(0,1):$ 
$
\frac{1}{\sqrt{n}}\sum_{i=1}^{\theta n} \down{\h}_i \eqP \phi\big(Q^{-1}(\theta)\big)\nn.
$
Thus, the objective function in \eqref{eq:Gor} converges in probability, point-wise on $\alpha$, to $\ell(\alpha^2/4)$ (cf. \eqref{eq:ell}). In fact, since the minimization in \eqref{eq:Gor} is over a compact set, uniform convergence holds and the minimum value converges to  $\min_{1\geq\alpha\geq\alphazero+\eps}\ell(\alpha^2/4)$.
Combining the above,
shows that for all $\eta>0$ the following event holds wpa 1:
\begin{align}\label{eq:LB}
\min_{\w\in\Sc(\alphazero)}  G(\w;\g,\h)\, \geq\, \min_{1\geq\alpha\geq\alphazero+\eps}  \ell\big(\alpha^2/4\big) - \eta.
\end{align}
Hence, from \eqref{eq:GMT_proof} the above statement holds with $G(\w;\g,\h)$ replaced by $F(\w;\z,\G)$.

Next, we obtain a simple upper bound on the RHS in \eqref{eq:toshow2}:
\begin{align}\label{eq:UB}
\min_{\w\in\{-2/\sqrt{n},0\}^n} F(\w) \leq F(\mathbf{0}) = \frac{\|\z\|_2}{\sqrt{n}},
\end{align}
which we combine with the fact that wpa 1 it holds ${\|\z\|_2}/{\sqrt{n}}\leq \sqrt{\delta}\,\sigma + \eta$.

Combining the two displays in \eqref{eq:LB} and \eqref{eq:UB}, we have shown that \eqref{eq:toshow2} holds as long as there exists $\eta>0$ such that
\begin{align} \label{eq:last2show}
\min_{1\geq\alpha\geq\alphazero+\eps}  \ell\big(\alpha^2/4\big) \geq \sqrt{\delta}\,\sigma + 3\eta.
\end{align}
At this point, recall that $\alphazero^2/4=\thetazero$ and the definition of $\theta_0$ as the largest solution to the equation $\ell(\theta)=\sqrt{\delta}\,\sigma$. By this definition and the fact that $\ell(\theta)$ is continuous and satisfies $\ell(1^-)>\sqrt{\delta}\,\sigma$ we have that $\ell(\theta)>\sqrt{\delta}$ for all $\theta>\thetazero$. Thus, there always exist $\eta(\eps)$ satisfying \eqref{eq:last2show} and the proof is complete.



%% file: replica.tex

\subsection{Gordon's prediction meets Tanaka}\label{sec:replica}

Inspecting the proof of Theorem \ref{thm:main} reveals two possible explanations for why the resulting upper bound might be loose. First, recall that we obtain a lower bound in the LHS of \eqref{eq:toshow2} via Gordon's inequality. As mentioned, in Remark \ref{rem:GMT} the inequality is not guaranteed to be tight in this instance. Second, recall that in upper bounding the RHS of \eqref{eq:toshow2} we use the crude bound \eqref{eq:UB}. Specifically, we upper bound the optimal cost $c_\star$ of the MAP in \eqref{eq:MAP} simply by the value of the objective function at a known feasible solution, namely $\x=\x_0$. 

In this section, we make the following leap of faith. We assume that $\inf_{\theta\in(0,1)}\ell(\theta)$ is an asymptotically \emph{tight} high-probability lower bound of $c_\star$, i.e., for all $\eta>0$ wpa 1:
\begin{align}\label{eq:assume}
\min_{\x\in\{\pm 1\}^n}\frac{1}{\sqrt{n}}\|\y-\A\x\|_2 \stackrel{?}{\leq} \inf_{\theta\in(0,1)}\ell(\theta) + \eta.
\end{align} 

Assuming \eqref{eq:assume} is true and repeating the arguments of Section \ref{sec:proof} leads to the following conclusion: the BER of the MAP detector is upper bounded by
$\thetastar=\arg\min_{\theta\in(0,1)}\ell(\theta).$ This can be also be expressed as the solution to the fixed-point equation $\ell'(\thetastar)=0$. Interestingly, this is shown in Proposition \ref{lem:ell_critical}(i) to be equivalent to \eqref{eq:replica}. In other words, under the assumption above, Gordon's prediction on the BER of the MAP detector coincides with the replica prediction under the RS ansatz. While it is known that the MAP detector exhibits replica symmetry breaking (RSB) behavior \cite{tanaka2001statistical}, we believe that our observation on a possible connection between Gordon's inequality and the replica symmetric prediction is worth exploring further. 
%

%

%% file: conclusion.tex
\section{Conclusion}
In this paper, we prove a simple yet highly non-trivial upper bound on the BER of the MAP detector in the case of BPSK signals and of equal-power condition. Theorem \ref{thm:main} naturally extends to allow for other constellation types (such as M-PAM) and power control and it also enjoys a non-asymptotic version. Perhaps more challenging, but certainly of interest, is the extension of our results to complex Gaussian channels. 
 Also, we wish to develop a deeper understanding of the connection between Gordon's inequality and the replica-symmetric prediction.

%% file: appendix.tex
\appendix
\section{Properties of $\ell(\theta)$}
Let $\ell:(0,1)\rightarrow\R$ and $\thetazero$ be defined as in Theorem \ref{thm:main}. The proofs of the propositions below are deferred to Appendices \ref{sec:proof1} and \ref{sec:proof2}.

\begin{propo}\label{lem:ell_critical}
The following statements are true:
\begin{enumerate}[(i)]
\item $\theta \in (0,1)$ is a critical point of $\ell$ if and only if it solves \eqref{eq:replica}.  All critical points belong in $(0,\frac{1}{2})$.
\item $\ell$ has either one or three critical points.
\item $\ell$ has a unique critical point if either one of the following two holds: $\delta>0.9251$ or $\sigma^2>0.1419$.
\end{enumerate}
\end{propo}

\begin{propo}\label{lem:ell_more}
If $\ell$ has a unique critical point (see Prop.~\ref{lem:ell_critical}(iii)), then the following are true:
\begin{enumerate}[(i)]
\item $\thetazero$ is the \emph{unique} solution of the equation $\ell(\theta)=\sigma\sqrt{\delta}$ in $(0,1)$.
\item The unique solution of \eqref{eq:replica} is the unique $\thetastar=\arg \min_{\theta} \ell(\theta)$.
\end{enumerate}

\noindent Moreover, if $\delta>0.9251$ it holds that:

\begin{enumerate}[(i)]
  \setcounter{enumi}{2}
\item The unique solution $\theta_*=\theta_*(\sigma)$ of \eqref{eq:replica} satisfies $\frac{\theta_*}{Q \big({\sqrt{\delta}}/{ \sigma }\big)} \rightarrow 1$, in the limit of $\sigma^2 \rightarrow 0$. 

\item For $\eta>0$, $\theta_0=\theta_0(\sigma)$ satisfies $\limsup_{\sigma \rightarrow 0} \frac{\theta_0}{Q \left(\frac{\sqrt{\delta}}{ \sigma}-\eta \right)} \leq 1.$
\end{enumerate}

\end{propo}

%% file: fullVer_appendix.tex
\section{Appendix}

\subsection{Proof of Proposition \ref{lem:ell_critical}}\label{sec:proof1}

Setting $u:=Q^{-1}(\theta)$, we will equivalently study the critical points of the function 
$$\ellt(u):=\ell(Q^{-1}(\theta)) = \sqrt{\delta}\sqrt{4Q(u)+\sigma^2} - 2\phi(u).$$
By simple algebra, 
$$
\ellt'(u) = -2Q'(u)\big( u - \sqrt{\delta}(4Q(u)+\sigma^2)^{-1/2} \big),
$$
where $Q'(u)=-\phi(u)$. Clearly, $\ellt'(u)<0$ for all $u\leq 0$. Thus, all critical points of $\ellt$ are in $(0,+\infty)$. Now, let us define
\begin{align}\label{eq:Fdef_TR}
F(u):=u^2(4Q(u)+\sigma^2).
\end{align}
Note that for $u>0$
\begin{align}\label{eq:Feqvell_RT}
\ellt'(u)=0 \Leftrightarrow F(u)=\delta.
 \end{align}
Using the transformation $\theta=Q(u)$ and simple algebra shows that the equation on the RHS of \eqref{eq:Feqvell_RT} is identical to \eqref{eq:replica}. This proves statement (i) of the proposition. 

Next, we study the function $F(u)$. It can be shown that
$
F'(u) = 2u\big( G(u) + \sigma^2 ),
$
where we define
\begin{align}\label{eq:G_TR}
G(u):=4Q(u)+2uQ'(u).
\end{align}
By differentiating $G$, setting the derivative equal to zero, and using the identity $Q''(u)=-uQ'(u)$, it can be shown that $G$ is decreasing in $[0,\sqrt{3}]$ and increasing in $[\sqrt{3},+\infty)$. In particular, 
\begin{align}\label{eq:Gmin_TR}
\forall u>0:\quad G(u) \geq G(\sqrt{3}) \approx -0.14183.
\end{align}
Thus, for $\sigma^2\geq 0.1419 > -G(\sqrt{3})$, it holds $F'(u)>0$ for all $u>0$. Thus, $F$ is increasing, which implies (cf. \eqref{eq:Feqvell_RT}) that $\ellt$ has a unique critical point. Moreover, for $\sigma^2\in\big[0,G(\sqrt{3})\big)$, the equation $F'(u)=0$ has two solution. Thus, $F$ has exactly two critical points, which we denote by $u_A$ and $u_B$, onwards. 

From the above properties of $F$, we conclude that $F$ is increasing in $[0,u_A]$, decreasing in $[u_A,u_B]$ and increasing in $[u_B,+\infty]$. Moreover, $u_A\leq \sqrt{3}$. Thus, for $\sigma^2\in\big[0,G(\sqrt{3})\big)$: $\ellt$ has three critical points if $\delta\in[F(u_A),F(u_B)]$ and one critical point, otherwise. This proves statement (ii) of the proposition.

Next, note that if $\delta>F(u_A)$ then $\ellt$ has a unique critical point. We will show that $F(u_A)\leq 0.9251$, thus establishing statement (iii) of the proposition. Using the fact that $G(u_A)+\sigma^2=0$, it follows that  $F(u_A) = -2u_A^3Q'(u_A)$. Now, setting $H(u)=-2u^3Q'(u)$, it can be readily shown by studying the derivative $H'(u)$, that 
\begin{align}\label{eq:F<_TR}
\max_{u> 0} -2u^3Q'(u) \leq H(\sqrt{3}) \approx 0.925082 < 0.9251,
\end{align} 
as desired. This concludes the proof of the proposition.

\subsection{Proof of Proposition \ref{lem:ell_more}}\label{sec:proof2}

\vspace{3pt}
(i) We can continuously extend $\ell$ to include the endpoints of the interval $[0,1]$. Note that $\ell(0)=\ell(0^+)=\sigma\sqrt{\delta}$. For the shake contradiction, suppose that there exists $0<\theta_1<\theta_0$ such that $\ell(\theta_1)=\sigma\sqrt{\delta}$. Then, by Rolle's theorem $\ell$ would have two distinct critical points, which contradicts the hypothesis.

\vspace{3pt}
(ii) For all $\theta \in (0,1)$, we have $\ell'(\theta) =\frac{2 \sqrt{\delta}}{\sqrt{4 \theta+\sigma^2}}-2 Q^{-1}(\theta)$. As for $\theta$ approaching $0$, $Q^{-1}(\theta)$ approaches $+\infty$, we can conclude that for $\theta$ sufficiently small, $\ell'(\theta)<0$. Similarly, as for $\theta$ approaching $1$, $Q^{-1}(\theta)$ approaches $-\infty$, we conclude that for $\theta$ sufficiently close to 1, $\ell'(\theta)>0$.  Given that $\ell'(\theta)=0$ has a unique solution $\theta_c$ we conclude that $\theta<\theta_c$ implies $\ell'(\theta)<0$ and $\theta>\theta_c$, $\ell'(\theta)>0$. In particular, $\theta_c$ is the global minimum of $\ell$, i.e. $\theta_c=\theta_*$.\\

(iii) We first establish that $\theta_* \rightarrow 0$, as $\sigma \rightarrow 0$.  To see this, consider by contradiction a limiting point, $\theta_L\in(0,1/2)$ of the function $\theta_*(\sigma)$. By \eqref{eq:replica} it must be true by taking limits, $\theta_L=Q(\sqrt\frac{{\delta}}{4\theta_L}).$ Thus, setting $u_L = Q^{-1}(\theta_L)$, it holds $F_0(u_L)=\delta$, where we defined the function $F_0(u):=4u^2Q(u)$ for $u\geq 0$. To conclude with a contradiction, we prove next that 
\begin{align}\label{eq:prop2ii}
\max_{u\geq0} F(u)<\delta.
\end{align}
Note that $F_0'(u)=2uG(u)$, where $G$ is defined in \eqref{eq:G_TR}. Let $u_{A0}$ be the solution of $G(u)=0$. From the proof of Proposition \ref{lem:ell_critical}, it is known that $u_{A0}$ is unique and maximizes $F_0(u)$. Moreover, it is easily seen that $F_0(u_{A0})=-2u_{A0}^2Q'(u_{A0})$. Thus, $\max_{u\geq0} F(u)\leq F_0(u_{A0}) \leq 0.92508\ldots$ .  where the last inequality was established in \eqref{eq:F<_TR}. This shows \eqref{eq:prop2ii}.


Now by mean value theorem, for some $\theta_T \in (0,\theta_*)$,
\begin{align*}
\theta_*-Q \left(\frac{\sqrt{\delta}}{ \sigma}\right)&=Q \left(\frac{\sqrt{\delta}}{\sqrt{\sigma^2+4\theta_*}}\right)-Q \left(\frac{\sqrt{\delta}}{\sqrt{\sigma^2}}\right)\\
&=\left[Q \left(\frac{\sqrt{\delta}}{\sqrt{\sigma^2+4\theta_T}}\right)\right]' \theta_*\\
&=\frac{2 \sqrt{\delta}}{(4 \theta_T+\sigma^2)^{\frac{3}{2}}}\,Q' \left(\frac{\sqrt{\delta}}{\sqrt{\sigma^2+4\theta_T}}\right)\theta_*\\
&=\frac{\sqrt{2} \sqrt{\delta}}{\sqrt{\pi}(4 \theta_T+\sigma^2)^{\frac{3}{2}}}e^{-\frac{\delta}{2(\sigma^2+4\theta_T)}}\theta_*.
\end{align*} 
Therefore,
\begin{equation} \label{limsup0} \limsup_{\sigma \rightarrow 0} \Big|1-\frac{Q \big(\frac{\sqrt{\delta}}{ \sigma}\big)}{\theta_*}\Big| \leq \limsup_{\sigma \rightarrow 0}\frac{\sqrt{2} \sqrt{\delta}}{\sqrt{\pi}(4 \theta_T+\sigma^2)^{\frac{3}{2}}}e^{-\frac{\delta}{2(\sigma^2+4\theta_T)}}.\end{equation} 
Since $0<\theta_T<\theta_* $ we know that $\theta_T$ also goes to zero as $\sigma$ goes to zero. In particular $\sigma^2+4\theta_T$ goes to zero and as $\delta$ is fixed, (\eqref{limsup0}) implies the desired result.\\

(vi) By statement (i) of the proposition, $\theta_0$ is the unique solution in $(0,1)$ of $\ell(\theta)=\sigma \sqrt{\delta}=\ell(0^+)$. 
Now we have $\theta_0>0$ satisfies $\ell(\theta_0)=\sigma \sqrt{\delta}$, or, 
\begin{equation} \label{main}
\sqrt{\delta}  \sqrt{4 \theta_0+\sigma^2}- 2\phi\big(Q^{-1}(\theta_0)\big) =\sqrt{\delta} \sigma .\end{equation}
We first prove that $\theta_0 \rightarrow 0$, as $\sigma \rightarrow 0$.  Indeed, if not, suppose $\theta_{N}>0$ is a positive limiting point of $\theta_0$ as $\sigma$ goes to zero. Then (\ref{main}) implies $L(\theta_N)=0$ for $L(\theta):=\sqrt{\delta}  \sqrt{4 \theta}-2\phi\big(Q^{-1}(\theta)\big)$. Since $L(0)=0$ by Rolle's theorem we have for some $\theta_L \in (0,1)$, $L'(\theta_L)=0$ which gives $\theta_L=Q(\sqrt\frac{{\delta}}{4\theta_L}).$ This leads to a contradiction, exactly as in the proof of statement (iii) above.
%

Now by (\ref{main}) by rearranging and $\phi$ the Gaussian density, we have  \begin{equation*} 
\sqrt{\delta} \left( \sqrt{4 \theta_0+\sigma^2}-\sigma\right)=\phi(Q^{-1}(\theta_0)).  \end{equation*}Taylor expansion around $0$ and the fact that $\theta_0=o(1)$ give $\phi(Q^{-1}(\theta_0))=2Q^{-1}(\theta_0)\theta_0+o(\theta_0).$ Hence we have \begin{equation*} 
\sqrt{\delta}\frac{1}{2 \theta_0} \left( \sqrt{4 \theta_0+\sigma^2}-\sigma\right)=Q^{-1}(\theta_0)+o(1) \end{equation*}or by simple algebra
\begin{equation*} 
\frac{2\sqrt{\delta}}{\sqrt{4 \theta_0+\sigma^2}+\sigma}=Q^{-1}(\theta_0)+o(1). \end{equation*}

Now since $\eta>0$, we conclude that for $\sigma$ sufficiently large, 
\begin{equation*} 
\frac{2\sqrt{\delta}}{\sqrt{4 \theta_0+\sigma^2}+\sigma}-\eta<Q^{-1}(\theta_0), \end{equation*}or
\begin{equation}\label{main2} 
\theta_0<Q\left(\frac{2\sqrt{\delta}}{\sqrt{4 \theta_0+\sigma^2}+\sigma}-\eta \right). \end{equation}
Finally, by (\ref{main2} and the mean value theorem we have that for some $\theta_T \in (0,\theta_0)$,
\begin{align*}
\theta_0-Q \left(\frac{\sqrt{\delta}}{ \sigma}-\eta \right)&<Q \left(\frac{2\sqrt{\delta}}{\sqrt{4\theta_0+\sigma^2}+\sigma}-\eta\right)-Q \left(\frac{\sqrt{\delta}}{\sqrt{\sigma^2}}-\eta \right)\\
&=\left[Q \left(\frac{2\sqrt{\delta}}{\sqrt{4\theta_T+\sigma^2}+\sigma}-\eta \right)\right]' \theta_0\\
&=\frac{2 \sqrt{\delta}}{(4 \theta_T+\sigma^2-\eta)^{\frac{3}{2}}}Q' \left(\frac{\sqrt{\delta}}{\sqrt{\sigma^2+4\theta_T}}-\eta\right)\theta_0\\
&=\frac{\sqrt{2} \sqrt{\delta}}{\sqrt{\pi}(4 \theta_T+\sigma^2-\eta)^{\frac{3}{2}}}e^{-\left(\frac{\sqrt{\delta}}{\sqrt{\sigma^2+4\theta_T}}-\eta\right)^2/2}\theta_0.
\end{align*} 
Therefore,
\begin{align} 
&1-\liminf_{\sigma \rightarrow 0} \frac{Q \left(\frac{\sqrt{\delta}}{ \sigma}-\eta\right)}{\theta_0} \notag \leq \\&
\qquad\limsup_{\sigma \rightarrow 0}\frac{\sqrt{2} \sqrt{\delta}}{\sqrt{\pi}(4 \theta_T+\sigma^2-\eta)^{\frac{3}{2}}}e^{-\left(\frac{\sqrt{\delta}}{\sqrt{\sigma^2+4\theta_T}}-\eta\right)^2/2}.\label{limsup2} 
\end{align} 
Since $0<\theta_T<\theta_0 $ we know that $\theta_T$ also goes to zero as $\sigma$ goes to zero. In particular $\sigma^2+4\theta_T$ goes to zero and as $\delta$ is fixed, (\ref{limsup2}) implies the desired result.\\


\subsection{Tanaka's equations}\label{sec:Tanaka}

For the reader's convenience we repeat Tanaka's fixed-point (FP) equations \cite[Eqn.~(43)]{Tanaka} using our notation. This is based on the following mapping between notation here and \cite{Tanaka}: $\delta\leftrightarrow \beta^{-1}$, $\sigma^2\leftrightarrow B_0^{-1}=\sigma_0^2/\beta$,  and $\BER\leftrightarrow (1-m)/2$.
\begin{align}
1 - 2\,\BER &= \int{\tanh({\sqrt{F}z+E})\,\phi(z) \mathrm{d}z} \nn\\
q &= \int{\tanh^2({\sqrt{F}z+E})\,\phi(z) \mathrm{d}z} \nn\\
 E&= \frac{\delta B}{1+B(1-q)}\nn\\
 F &= \frac{\delta B^2(\sig^2+4\,\BER+q-1)}{(1+B(1-q))^2},\nn
\end{align}
where for the MAP decoder one needs to solve the equations above for $B\rightarrow\infty$. In this case,
\begin{align}
1 - 2\,\BER &= \int{\tanh({\sqrt{F}z+E})\,\phi(z) \mathrm{d}z} \nn\\
q &= \int{\tanh^2({\sqrt{F}z+E})\,\phi(z) \mathrm{d}z} \nn\\
 E&= \frac{\delta}{(1-q)}\nn\\
 F &= \frac{\delta (\sig^2+4\BER+q-1)}{(1-q)^2} \nn\\
 &\qquad \Leftrightarrow \sqrt{F} = E\sqrt{\frac{\sig^2+4\,\BER+q-1}{\delta}},\label{eq:Tanaka_original}
\end{align}
It is empirically observed in \cite[Sec.~V.A]{Tanaka} that for $\delta>1/1.08$, the FP equations \eqref{eq:Tanaka_original}  have a unique solution. Now, consider \eqref{eq:replica}. Let $\delta>1.46$ such that \eqref{eq:replica} has a unique solution $\thetastar$. Then, it is not hard to see that the quadruple 
$\big(\,m\rightarrow1-2\thetastar,\, q\rightarrow 1, \, E\rightarrow+\infty, F\rightarrow+\infty \,\big)$ satisfies \eqref{eq:Tanaka_original}. Indeed, denoting 
$$c:=c(\BER) = \sqrt{\frac{\sigma^2+4\,\BER}{\delta}},$$ note that for $q\rightarrow 1$: $\sqrt{F}=c\,E$. Next, using that $$\tanh\big(E({c\,z+1})\big){\rightarrow}\ind{z\geq -c^{-1}}-\ind{z\leq -c^{-1}}$$
as $E\rightarrow+\infty$, shows that the second equation in  \eqref{eq:Tanaka_original} is consistent with $q\rightarrow 1$. Finally, the first equation becomes
\begin{align*}
1-2\,\BER &= \Pr(z\geq -c^{-1}) - \Pr(z\leq -c^{-1})\\ &= 1 - 2\Pr(z\leq -c^{-1}) = 1 - 2 Q(c^{-1}),
\end{align*}
which agrees with \eqref{eq:replica}, as desired.